# Vibrational features in inelastic electron tunneling spectra

Kamil Walczak

School of Electrical & Computer Engineering, University of Virginia
351 McCormick Road, Charlottesville, Virginia 22904, USA

A theoretical analysis of inelastic electron tunneling spectroscopy (IETS) experiments conducted on molecular junctions is presented, where the second derivative of the current with respect to voltage is usually plotted as a function of applied bias. Within the nonperturbative computational scheme, adequate for arbitrary parameters of the model, we consider the virtual conduction process in the off-resonance region. Here we study the influence of few crucial factors on the IETS spectra: the strength of the vibronic coupling, the phonon energy, and the device working temperature. It was also shown that weak asymmetry in the IETS signal with respect to bias polarity is obtained as a result of strongly asymmetric connection with the electrodes.

Key words: inelastic electron tunneling spectroscopy (IETS), phonon-assisted tunneling, inelastic transport, vibronic coupling, molecular electronics

PACS numbers:  85.65.+h, 73.90.+f, 71.38.-k

## I. Introduction

Inelastic electron tunneling spectroscopy (IETS) is a powerful experimental tool for identifying and characterizing molecular species within the conduction area [1-10]. Standard ac modulation techniques, along with two lock-in amplifiers, are utilized to measure current-voltage ($J-V$) characteristics as well as the first and the second harmonic signals (proportional to $dJ/dV$ and $d^2J/dV^2$, respectively). This method provides information on the strength of the vibronic coupling between the charge carriers and nuclear motions of the molecules. The IETS experiment can also be helpful in identifying the geometrical structures of molecules and molecule-metal contacts, since junctions with different geometries disclose very different spectral profiles [7,8]. The measured spectra show well-resolved vibronic features corresponding to certain vibrational normal modes of the molecules. It is also well-known that the IETS spectra are very sensitive also to few other factors, such as: the strength of the molecule-metal bonding, and the intramolecular conformational changes.

In the literature, we can distinguish two different approaches to the problem of inelastic transport. One of them is based on low order perturbative treatment, where the tunneling current is computed in the lowest order with respect to the strength of the electron-phonon interaction [11-18]. This approach is usually used to explain the IETS spectra, since the vibronic coupling in those experiments is assumed to be relatively small. However, for low-frequency vibrations, the parameter of the electron-phonon interaction is of the same order of magnitude or larger ($\sim 0.3 eV$) and the perturbative treatment is inadequate. Besides, this approximation is not fully consistent with the boundary restrictions imposed by the Pauli exclusion principle. The second mentioned method is associated with nonperturbative treatment, where the many-body electron-phonon interaction problem is transformed into a one-body many-channel scattering problem within the so-called mapping technique [19-25]. The main advantage of this approach is the fact that it does not involve any restrictions on the model parameters – we can use it even for the case of extremely strong vibronic coupling.

The main purpose of this work is to use nonperturbative method, based on Green's function theory, to study the shape of transport characteristics observed in the IETS experiments in relation to three crucial factors: the vibronic coupling, the phonon energy and the temperature of the system under investigation. Molecular vibrations are modeled as dispersionless phonon excitations which can locally interact with conduction electrons. At the threshold voltage, where the potential bias between the two electrodes exceeds the phonon energy, the tunneling electron can exchange energy with this mode and the inelastic signal is (usually) observed as a conductance step or a peak in the $d^2J/dV^2$ versus voltage $V$.

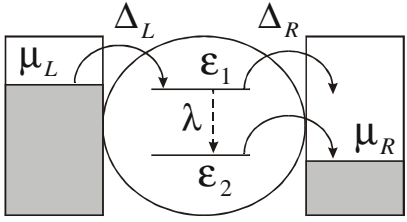

**Figure 1:** A schematic representation of the phonon-assisted tunneling process in the resonance case.

## II. Model

Here we consider the arbitrary vibronic coupling case in the near-resonance region, although we restrict ourselves to one-phonon case (the higher excitations can be easily incorporated into our scheme [19]). When the bias voltage is larger than the phonon excitation energy, the electron incoming from the source reservoir can loose part of its energy (emit one phonon) on the molecule and then outgo into the drain reservoir (inelastic contribution) or can outgo to the drain without exchanging energy with nuclear degrees of freedom (elastic part). If we restrict ourselves to the simplest possible case of one electronic level $\varepsilon_1$ coupled to a single phonon mode $\Omega$ (see Fig.1), the Hamiltonian can be written in the convenient matrix form:

$$H_{eff} = \begin{bmatrix} \varepsilon_1 - i(\Delta_L + \Delta_R) & \lambda \\ \lambda & \varepsilon_2 - i\Delta_R \end{bmatrix}, \quad (1)$$

where $\lambda$ denotes the strength of the electron-phonon coupling, while $\varepsilon_2 = \varepsilon_1 - \Omega$ is virtual energy level participating in the conduction process. The corresponding Green function ($G(\varepsilon) = [\varepsilon J - H_{eff}]^{-1}$, where $J$ is the identity matrix of Hamiltonian dimensions):

$$G(\varepsilon) = \begin{bmatrix} \frac{1}{Q}[\varepsilon - \varepsilon_2 + i\Delta_R] & \frac{\lambda}{Q} \\ \frac{\lambda}{Q} & \frac{1}{Q}[\varepsilon - \varepsilon_1 + i(\Delta_L + \Delta_R)] \end{bmatrix}, \quad (2)$$

where the denominator:

$$Q = [\varepsilon - \varepsilon_1 + i(\Delta_L + \Delta_R)][\varepsilon - \varepsilon_2 + i\Delta_R] - \lambda^2. \quad (3)$$

Here $\Delta_L$ and $\Delta_R$ are the so-called escape rates into the left (source) and right (drain) electrodes taken in the wide-band approximation. The total transmission can be written as the sum of elastic and inelastic parts:

$$T_{tot}(\varepsilon) = T_{el}(\varepsilon) + T_{inel}(\varepsilon), \quad (4)$$

$$T_{el}(\varepsilon) = 4\Delta_L \Delta_R \mid G_{11}(\varepsilon) \mid^2 = \frac{4\Delta_L \Delta_R}{M} \left[ (\varepsilon - \varepsilon_2)^2 + \Delta_R^2 \right], \tag{5}$$

$$T_{inel}(\varepsilon) = 4\Delta_L \Delta_R \mid G_{12}(\varepsilon) \mid^2 = \frac{4\Delta_L \Delta_R \lambda^2}{M}, \tag{6}$$

where the denominator is given through the relation:

$$M = \left[ (\varepsilon - \varepsilon_1)(\varepsilon - \varepsilon_2) - \Delta_R(\Delta_L + \Delta_R) - \lambda^2 \right]^2 + \left[ \Delta_R(\varepsilon - \varepsilon_1) + (\Delta_L + \Delta_R)(\varepsilon - \varepsilon_2) \right]^2 \tag{7}$$

It should be noted that inelastic term of the transmission is directly proportional to $\lambda^2$, so in the case of $\lambda = 0$ the contribution from the virtual conduction process vanishes $T_{inel}(\varepsilon) = 0$, as expected. Since the transmission functions are divided into the elastic and inelastic contributions, also the current can be written as the sum of particular contributions [19-25]:

$$J_{tot} = J_{el} + J_{inel}, \tag{8}$$

$$J_{el} = \frac{2e}{h} \int_{-\infty}^{+\infty} T_{el}(\varepsilon) [f_L(\varepsilon) - f_R(\varepsilon)] d\varepsilon, \tag{9}$$

$$J_{inel} = \frac{2e}{h} \int_{-\infty}^{+\infty} T_{inel}(\varepsilon) [f_L(\varepsilon)[1 - f_R(\varepsilon - \Omega)] - f_R(\varepsilon)[1 - f_L(\varepsilon - \Omega)]] d\varepsilon. \tag{10}$$

Here $f_L(\varepsilon)$ and $f_R(\varepsilon)$ are Fermi distribution functions defined with the help of the following electrochemical potentials:

$$\mu_L = \varepsilon_F + \eta eV, \tag{11}$$

$$\mu_R = \varepsilon_F - (1 - \eta)eV, \tag{12}$$

$\varepsilon_F$ is equilibrium Fermi energy, $\eta = \Delta_L /(\Delta_L + \Delta_R)$ is the so-called voltage division factor [26], while $V = (\mu_L - \mu_R)/e$ is applied voltage. Eqs. (6) and (10) will be used in the next section to analyze the IETS spectra.

Now let us concentrate for a moment on the weak vibronic coupling case in the near-resonance region. To give some simple analytic results, let us choose the crucial coefficient in our analysis as $g = (\lambda/\Omega)^2$. Expanding the particular elements of the Green function into the Taylor series with respect to $g \ll 1$, we can simplify them to the expressions [22]:

$$G_{11}(\varepsilon) \cong \frac{1}{\varepsilon - \bar{\varepsilon}_1 + i(\Delta_L + \Delta_R)}, \tag{13}$$

$$G_{12}(\varepsilon) \cong \frac{\sqrt{g}}{\varepsilon - \bar{\varepsilon}_2 + i(g\Delta_L + \Delta_R)}, \tag{14}$$

where the energy of the molecular level shifted by polaron correction is defined as: $\bar{\varepsilon}_1 = \varepsilon_1 - \lambda^2/\Omega$. The first element of that Green function describes the elastic part of the transmission (process without the phonon emission during the tunneling), while the second element contributes to the total transmission as its inelastic part (the molecule is excited to vibrations before the electron reaches drain). For the case of $g \ll 1$, the elastic and inelastic parts of the total transmission are given through the approximate relations:

$$T_{el}(\varepsilon) = 4\Delta_L \Delta_R |G_{11}(\varepsilon)|^2 \cong \frac{4\Delta_L \Delta_R}{[\varepsilon - \bar{\varepsilon}_1]^2 + [\Delta_L + \Delta_R]^2}, \tag{15}$$

$$T_{inel}(\varepsilon) = 4\Delta_L \Delta_R |G_{12}(\varepsilon)|^2 \cong \frac{4g\Delta_L \Delta_R}{[\varepsilon - \bar{\varepsilon}_2]^2 + [g\Delta_L + \Delta_R]^2}. \tag{16}$$

Now let us consider the special condition $\varepsilon = \bar{\varepsilon}_1$ in the symmetric coupling case: $\Delta_L = \Delta_R = \Delta$. Since inelastic component of the transmission function is directly proportional to the $g$-parameter, for the case of $g \to 0$ we have $T_{inel} = 0$ and $T_{el} = T_{tot} = 1$ (perfect one-channel tunneling). For the special case of $g = 1$ the transmission function looks like two Lorentzian peaks with maxima equal to one for energies: $\varepsilon = \bar{\varepsilon}_1$ and $\varepsilon = \bar{\varepsilon}_2$ (two-channel tunneling). Moreover, since there is no direct connection between the virtual conduction channel $\varepsilon_2$ and the left electrode (the escape rate is equal to zero), the orbital-metal coupling $\Delta_L$ in Eq.(16) is reduced by the $g$-factor (since $g \ll 1$). We can also easily find an analytical expression for the low-temperature elastic part of the current:

$$J_{el} \cong \frac{8e}{h} \frac{\Delta_L \Delta_R}{\Delta_L + \Delta_R} \left[ \arctan\left(\frac{\bar{\varepsilon}_1 - \mu_R}{\Delta_L + \Delta_R}\right) - \arctan\left(\frac{\bar{\varepsilon}_1 - \mu_L}{\Delta_L + \Delta_R}\right) \right]. \tag{17}$$

So far, we have analyzed phonon-assisted tunneling in near-resonance region. Anyway, we know from the experiments that the virtual conduction process is realized even in the off-resonance case, every time the voltage energy coincides with the energy of the vibrational mode [1-10]. In the deep tunneling phenomenon, where $\varepsilon \ll \bar{\varepsilon}_2$, the Lorentzian tail of the transmission function is so smooth that it becomes practically dispersionless:

$$T_{inel} = g \frac{4\Delta_L \Delta_R}{\bar{\varepsilon}_2^2 + [g\Delta_L + \Delta_R]^2} \equiv gD_0, \tag{18}$$

Now the formula for the inelastic current given by Eq.(10) is simplified to the expression:

$$J_{inel} \cong \frac{2e}{h} gD_0 \int_{-\infty}^{+\infty} [f_L(\varepsilon)[1 - f_R(\varepsilon - \Omega)] - f_R(\varepsilon)[1 - f_L(\varepsilon - \Omega)]] d\varepsilon, \tag{19}$$

Further simplification stems from assumption of low-temperature limit, where the Fermi distribution can be replaced by the Heaviside step function, while its derivative is simply Dirac delta function. Let us also assume that the bias voltage is dropped entirely on the drain electrode. Now we can calculate the conductance as the first derivative of the current with respect to voltage:

$$\frac{\partial J_{inel}}{\partial V} \cong \frac{2e^2}{h} gD_0 [\theta(eV - \Omega) - [1 + \theta(eV + \Omega)]]. \tag{20}$$

The above Eq.(20) predicts the step-like character of the conductance-voltage dependence, where the jumps appear every time the bias voltage is in resonance with molecular vibration: $eV = \pm\Omega$. This result can be easily transformed into the peaks observed for the second derivative of the current with respect to voltage:

$$\frac{\partial^2 J_{inel}}{\partial V^2} \cong \frac{2e^3}{h} g D_0 [\delta(eV - \Omega) - \delta(eV + \Omega)]. \qquad (21)$$

Such analytical proves are in good qualitative agreement with IETS experimental data.

### III. Numerical results and discussion

Now, let us proceed to consider just one electronic level $\varepsilon_1$ coupled to a single phonon mode with energy $\Omega$ (the coupling strength is denoted as $\lambda$) which is connected to two broad-band metals (descried with the help of the dispersionless parameters $\Delta_L$ and $\Delta_R$). This is a test case simple enough to analyze the essential physics of inelastic transport problem in detail. Besides, generalization to multilevel system with many different phonons can be obtained straightforwardly. Here we choose the following model parameters (energies are given in $eV$): $\varepsilon_1 = 5$, $\varepsilon_F = 0$ (off-resonance limit), $\Delta_L = \Delta_R = 0.1$ (relatively weak connections with the electrodes). The electron-phonon coupling coefficient $\lambda$ can be estimated in molecular systems from reorganization energies: $E_{reorg} \cong \lambda^2 / \Omega$, inferred from electron-transfer rate studies in similar environments. Since observed values of $E_{reorg}$ are $0.1 - 1 eV$ and $\Omega \sim 0.1 eV$ (as registered in experiments), the magnitude of $\lambda$ is placed in the range of $0.1 - 0.3 eV$. Temperature in the experiments can be changed in the wide range, here we choose $4 - 300K$.

The voltage-dependence of the second derivative of the inelastic current with respect to voltage for three different values of the $\lambda$-parameter is illustrated in Fig.2. Since this graph is antisymmetrical with respect to bias polarity, thus for clarity we show only the positive bias region in the spectrum. The peak in the considered characteristic is higher for higher values of the vibronic coupling, while its line width is not changing. In fact, the peak of the $d^2 J_{inel} / dV^2$ spectrum scales as $\lambda^2$ (as shown in Fig.3 and predicted by Eq.(21)), while the slope of that straight line decreases with the increase of temperature. This observation is in good agreement with experimental results, where for higher temperatures the peaks in the IETS spectra reach lower values [9].

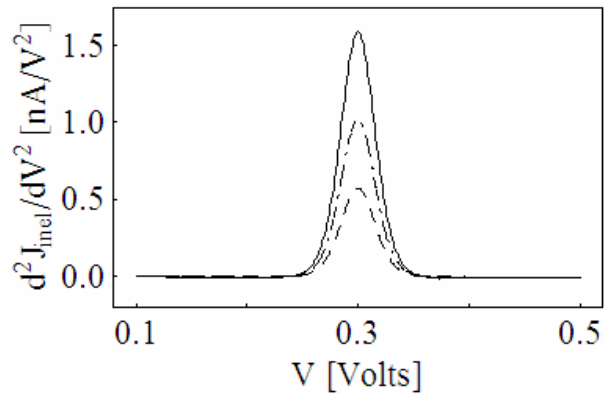 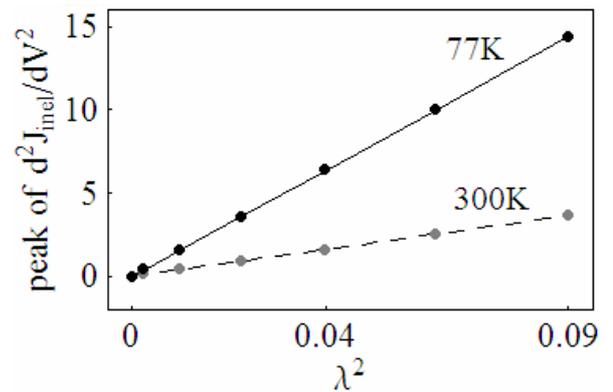

**Figure 2:** $d^2 J_{inel} / dV^2$ as a function of bias voltage at $\theta = 77K$ for three different vibronic coefficients: $\lambda = 0.06$ (dashed line), $\lambda = 0.08$ (dashed-dotted line), $\lambda = 0.10$ (solid line). Parameters (in $eV$): $\varepsilon_1 = 5$, $\varepsilon_F = 0$, $\Delta_L = \Delta_R = 0.1$, $\Omega = 0.3$.

**Figure 3:** Peak of $d^2 J_{inel} / dV^2$ (in $nA/V^2$) as a function of $\lambda^2$ (in $(eV)^2$) for two different temperatures: $\theta = 77K$ (black circles, solid line) and $\theta = 300K$ (grey circles, dashed line). The other parameters are the same as in Fig.2.

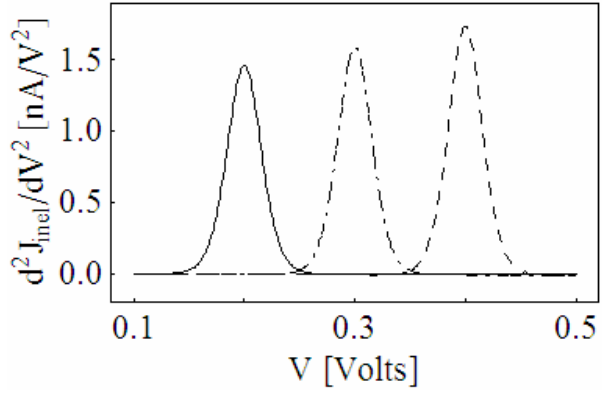 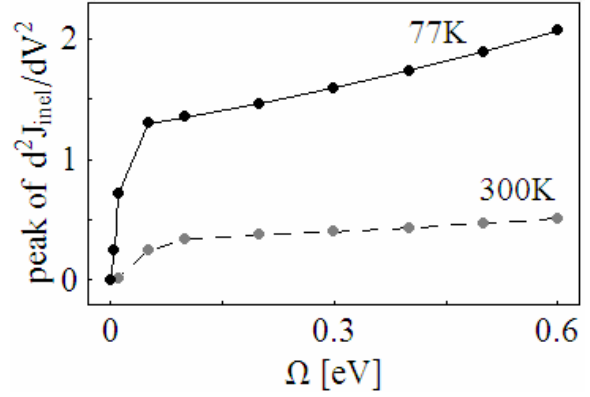

**Figure 4:** $d^2 J_{inel}/dV^2$ as a function of bias voltage at $\theta = 77K$ for three different phonon energies: $\Omega = 0.2$ (solid line), $\Omega = 0.3$ (dashed-dotted line), $\Omega = 0.4$ (dashed line). Parameters (in $eV$): $\varepsilon_1 = 5$, $\varepsilon_F = 0$, $\Delta_L = \Delta_R = 0.1$, $\lambda = 0.1$.

**Figure 5:** Peak of $d^2 J_{inel}/dV^2$ (in $nA/V^2$) as a function of $\Omega$ (in $eV$) for two different temperatures: $\theta = 77K$ (black circles, solid line) and $\theta = 300K$ (grey circles, dashed line). The other parameters are the same as in Fig.4.

The voltage-dependence of $d^2 J_{inel}/dV^2$ for three different values of the phonon energies is presented in Fig.4. It is obvious that the location of the peak of the considered dependence is determined by $\Omega$, as predicted by Eq.(21). However, for higher phonon energies we can observe an increase of the amplitude in IETS signal, as documented in Fig.5. The biggest changes are visible for $\Omega < 0.1 eV$, while for higher values of the phonon energy the height of the peak of $d^2 J_{inel}/dV^2$ spectrum is a smooth function of bias voltage. Low phonon frequencies observed in inelastic tunneling experiments are usually related to the stretching modes of the bonding electrode-molecule atoms [8-10]. Since the amplitudes in IETS spectra for $V < 0.1V$ are usually comparable with the other amplitudes, we conclude that the vibronic coupling in this region is strong.

The voltage-dependence of $d^2 J_{inel}/dV^2$ for three different temperatures is plotted in Fig.6. Here we can distinguish two different effects associated with the increase of the temperature: reduction of the peak amplitude and broadening of the linewidth, as shown in Figs.7 and 8, respectively. The peak of the $d^2 J_{inel}/dV^2$ spectrum is a strongly nonlinear function of temperature $\theta$, while its linewidth is changing linearly with temperature $\theta$. The temperature changes in our model are hidden into the exponential temperature-dependence of the Fermi distributions.

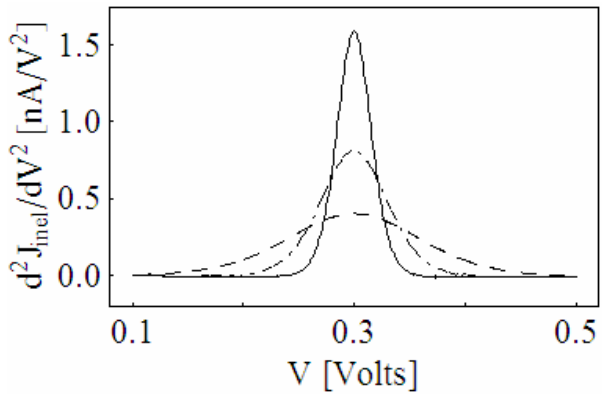

**Figure 6:** $d^2 J_{inel}/dV^2$ as a function of bias voltage for three different temperatures: $\theta = 77K$ (solid line), $\theta = 150K$ (dashed-dotted line), $\theta = 300K$ (dashed line). Parameters (in $eV$): $\varepsilon_1 = 5$, $\varepsilon_F = 0$, $\Delta_L = \Delta_R = 0.1$, $\Omega = 0.3$, $\lambda = 0.1$.

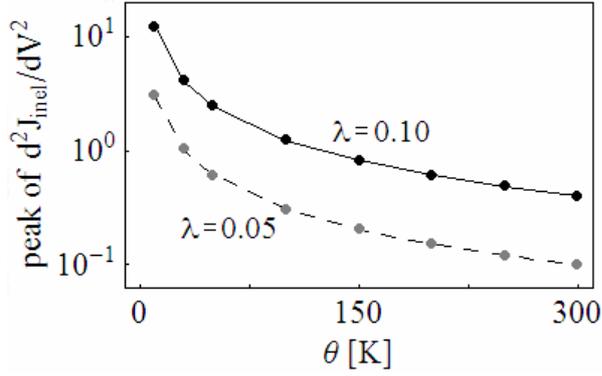
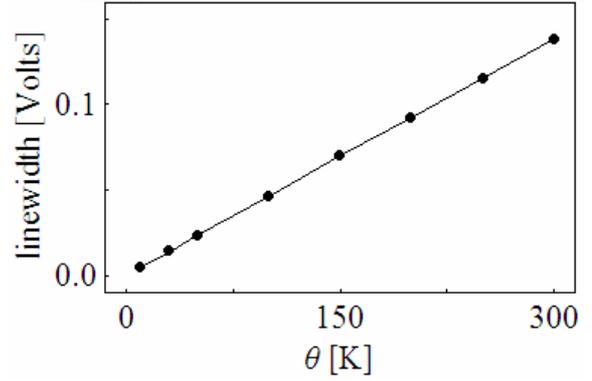

**Figure 7:** Peak of $d^2J_{inel}/dV^2$ (in $nA/V^2$) as a function of temperature $\theta$ (in $K$) for two different vibronic couplings: $\lambda = 0.10 eV$ (black circles, solid line) and $\lambda = 0.05 eV$ (grey circles, dashed line). The other parameters of the model are the same as in Fig.6. This picture is plotted in the semi-logarithmic scale.

**Figure 8:** Linewidth (given in Volts) of $d^2J_{inel}/dV^2$ vs. voltage curve as a function of temperature $\theta$ (in $K$) for $\lambda = 0.10 eV$ (this picture remains unchanged for a wide range of vibronic coupling coefficient). The other parameters are the same as in Fig.6.

The formalism presented in the previous section can be easily generalized to multiorbital system with many different phonon modes. Just to give a simple example, let us take a look at the results obtained for one energy level with two vibrational modes, as shown in Fig.9. For symmetrical molecule-to-electrodes connection case, we observe the inelastic spectrum in good quantitative agreement with some experimental data [10] thanks to the proper choice of the model parameters. The high-level background registered in the IETS experiment can be a consequence of the other molecular excitations or the elastic contribution, which is neglected in this paper, but could be essential when the molecular energy level is closer to the Fermi energy. Interestingly, for strongly asymmetrical connection case, a weak asymmetry in the IETS spectrum with respect to bias inversion is obtained and sometimes observed in the experiments [10]. In our calculations we used dispersionless escape rates, while the difference in the energy dependence of two contact self-energies can also contributes to the mentioned asymmetry [17].

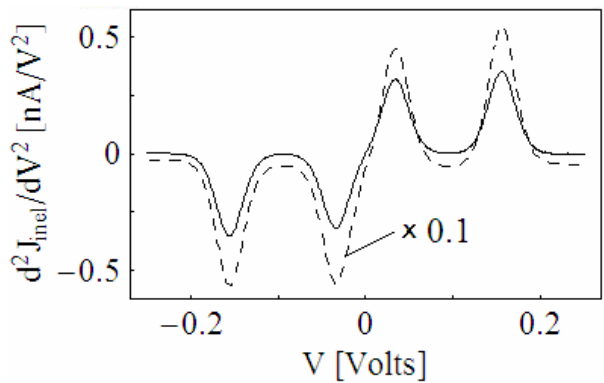

**Figure 9:** $d^2J_{inel}/dV^2$ as a function of bias voltage at $\theta = 77K$ for symmetric ($\Delta_L = \Delta_R = 0.1$: solid line) and strongly asymmetric ($\Delta_L = 100\Delta_R = 0.4$: dashed line) coupling with the electrodes. Parameters (in $eV$): $\varepsilon_1 = 5$, $\varepsilon_F = 0$, $\Omega_1 = 0.034$, $\Omega_2 = 0.156$, $\lambda_1 = \lambda_2 = 0.05$.

## IV. Concluding remarks

In summary, we have used the nonperturbative computational scheme (suitable for arbitrary model parameters) in order to study the influence of few crucial factors on the characteristics of the IETS experiments. The numerical calculations successfully reproduce the typical features observed by inelastic electron tunneling spectroscopy. In particular, it was shown that the height of the peak in the spectrum scales as square of the vibronic coupling coefficient, while it is strongly nonlinear function of phonon energy and temperature (for fixed other parameters). Besides, its linewidth scales linearly with temperature, while it does not change significantly upon the variations of the other analyzed parameters. Weak asymmetry of the IETS spectra with respect to bias inversion was also observed for the case of strongly asymmetric coupling with the electrodes.

It should be noted that our approach to inelastic transport may be easily applied for more realistic models of molecular energy structure with many different phonons, where the elastic contribution to the IETS spectrum is also included. In fact, the correction associated with this elastic part may be negative, and for some energetic parameters may even outweigh the positive contribution of the inelastic part resulting in the dip (instead of a peak) in the IETS response [27,28]. Our results confirm experimental data of a peak-width broadening effect as a function of temperature at a fixed (undetermined) modulation voltage. However, an interesting unsolved problem is to find theoretical explanation of the fact that IETS peaks and linewidths increasing with the increase of modulation voltage at a fixed temperature [29]. Actually, both mentioned effects constitute the test for the observed spectra that they are indeed valid IETS data.

## Acknowledgement

The author is very grateful to A.W. Ghosh, J.C. Bean, L.R. Harriott, K.A. Williams, L. Pu and S. Vasudevan for many valuable discussions.

## References

[1] R.C. Jaklevic, J. Lambe, Phys. Rev. Lett. **17**, 1139 (1966).
[2] K.W. Hipps, U.J. Mazur, J. Phys. Chem. **97**, 7803 (1993).
[3] B.C. Stipe, M.A. Rezaei, W. Ho, Science **280**, 1732 (1998).
[4] H. Park, J. Park, A.K.L. Lim, E.H. Anderson A.P. Alivisatos, P.L. McEuen, Nature **407**, 57 (2000).
[5] J.R. Hahn, H.J. Lee, W. Ho, Phys. Rev. Lett. **85**, 1914 (2000).
[6] R.H.M. Smit, Y. Noat, C. Untiedt, N.D. Lang, M.C.V. Hemert, J.M.V. Ruitenbeek, Nature **419**, 906 (2002).
[7] N.B. Zhitenev, H. Meng, Z. Bao, Phys. Rev. Lett. **88**, 226801 (2002).
[8] J.G. Kushmerick, J. Lazorcik, C.H. Patterson, R. Shashidhar, Nano Lett. **4**, 639 (2004).
[9] W. Wang, T. Lee, I. Kretzschmar, M.A. Reed, Nano Lett. **4**, 643 (2004).
[10] A.-S. Hallbäck, N. Oncel, J. Huskens, H.J.W. Zandvliet, B. Poelsema, Nano Lett. **4**, 2393 (2004).
[11] J. Kirtley, D.J. Scalapino, P.K. Hansma, Phys. Rev. B **14**, 3177 (1976).
[12] S. Datta, Nanotechnology **15**, S433 (2004).

[13] A. Troisi, M.A. Ratner, A. Nitzan, J. Chem. Phys. **118**, 6072 (2003).
[14] A. Troisi, M.A. Ratner, Phys. Rev. B **72**, 033408 (2005).
[15] A. Troisi, M.A. Ratner, Nano Lett. **6**, 1784 (2006).
[16] M. Galperin, M.A. Ratner, A. Nitzan, Nano Lett. **4**, 1605 (2004).
[17] M. Galperin, A. Nitzan, M.A. Ratner, D.R. Stewart, J. Phys. Chem. B **109**, 8519 (2005).
[18] M. Galperin, A. Nitzan, M.A. Ratner, Phys. Rev. B **73**, 045314 (2006).
[19] J. Bonča, S.A. Trugman, Phys. Rev. Lett. **75**, 2566 (1995).
[20] K. Haule, J. Bonča, Phys. Rev. B **59**, 13087 (1999).
[21] E.G. Emberly, G. Kirczenow, Phys. Rev. B **61**, 5740 (2000).
[22] L.E.F. Foa Torres, H.M. Pastawski, S.S. Makler, Phys. Rev. B **64**, 193304 (2001).
[23] M. Čížek, M. Thoss, W. Domcke, Phys. Rev. B **70**, 125406 (2004).
[24] K. Walczak, Physica E **33**, 110 (2006).
[25] K. Walczak, J. Magn. Magn. Mater. **305**, 475 (2006).
[26] W. Tian, S. Datta, S. Hong, R.G. Reifenberger, J.I. Henderson, C.P. Kubiak, Physica E **1**, 304 (1997).
[27] A. Bayman, P. Hansma, W.C. Kaska, Phys. Rev. B **24**, 2449 (1981).
[28] B.N.J. Persson, A. Baratoff, Phys. Rev. Lett. **59**, 339 (1987).
[29] W. Wang, T. Lee, M.A. Reed, J. Phys. Chem. B **108**, 18398 (2004).